# Vectorial nonlinear optics: Type-II second harmonic generation driven by spin-orbit coupled fields

Hai-Jun Wu,[1] Hao-Yan Yang,[1] Carmelo Rosales-Guzmán[1], Wei Gao,[1] Bao-Sen Shi,[1,2] and Zhi-Han Zhu[1,*]

[1] *Wang Da-Heng Collaborative Innovation Center, Heilongjiang Provincial Key Laboratory of Quantum Manipulation & Control, Harbin University of Science and Technology, Harbin 150080, China*

[2] *CAS Key Laboratory of Quantum Information, University of Science and Technology of China, Hefei, 230026, China*

Vectorial nonlinear optics refers to the investigation of optical processes whose nonlinear polarization (NP) undergoes spin-orbit coupling (SOC) interactions where, in general, the driving light field or the new field generated by the interaction containing the SOC property. To contribute to fundamental knowledge in this domain, we examine the type-II second harmonic generation (SHG) induced by vectorial laser modes. First, we provide a general theory to analyze the vectorial SHG process. Second, by using two typical vector modes as examples, we show how the SOC of the pump field dictates nonlinear interaction. Finally, we corroborate our theoretical predictions through experiments to confirm the crucial role of the SOC in nonlinear interactions. These results enhance our fundamental understanding of SOC-mediated nonlinear optics and lay the foundation for further fundamental studies as well as possible applications.

## I. INTRODUCTION

Since Maiman announced the first operative laser in 1960 based on Townes and Schawlow's theory [1], the unprecedented brightness of the coherent light obtained made it possible to generate strong light–matter interactions that enable high-order optical response from the materials, or rather nonlinear polarization (NP), to reach an observable level. Only a year later, Franken et al. observed the first laser radiation-induced nonlinear effect—i.e., second harmonic generation (SHG)—that prompted research on nonlinear optics [2]. Owing to its fundamental aspects and potential applications, nonlinear optics has rapidly emerged as an important subfield of modern optics and photonics [3-15]. Today, driven by continued advances in laser techniques over the past several decades, the longitudinal dimension of this field has expanded into ultrashort time and ultrahigh-intensity regions [16-19]. By contrast, in the transverse dimension, both from the theoretical and applied aspects, research has focused on Gaussian ($TEM_{00}$) beams with scalar states of polarization (SoPs). This can be attributed to a fact that although research on laser fields with spatially variant SoPs, i.e., the so-called vector mode, traces back to the 1970s [20, 21], most commercial laser resonators are designed to generate $TEM_{00}$ beams or pulses with a linear polarization for highly efficient generation.

More recently, progress in the physics behind spatially variant SoPs, i.e., photonic spin-orbit coupling (SOC) interactions [22-25], has significantly enhanced our capability to shape and control light–matter interactions. These highly customizable interactions have promoted a series of advances in photonics, ranging from super-resolution microscopy, laser trapping, and optical metrology, to fundamental physics [26-31]. Remarkably, the SOC-mediated light–matter interactions can provide an additional interface to tailor nonlinear optical processes that can significantly extend our understanding of nonlinear optics. Specifically, as the origin of nonlinear interactions, NP, i.e., the electric dipole moment in nonlinear media driven by laser fields, relies heavily on both the SoP and the spatial structure (i.e., phase and intensity profiles) of the applied fields. In the vectorial case, NPs created by vector modes thus possess significantly SoP-dependent spatial structures. These SOC-mediated NPs, on the one hand, introduce a host of vector nonlinear optical phenomena and rich functionalities that have no analog in scalar interactions, and, on the other hand, provide a promising way to generate, control, and characterize structured light. As a consequence, this emerging field has garnered growing interest in research and revived work on fundamental nonlinear optics [32-44].

The central premise of exploring this emerging field is to develop a general theoretical toolkit that can completely describe and analyze vectorial nonlinear optical processes, and yet it is a missing fundamental component. In light of this situation, we conduct a series of fundamental studies—the vectorial nonlinear optics series—involving typical second- and third-order nonlinear optical processes. Specifically, for a second-order interaction, owing to the fixed polarization dependence of the nonlinear medium, the reference frame of the polarization of the medium and the SOC state of the applied field codetermine the spatial structure of the excited NP. For a third-order interaction, the medium has more freedom to generate NPs, and thus more extraordinary phenomena can be expected. Remarkably, the boundary between classical and quantum optics is blurring. The quantum attribute of "non-separability" can now be exploited in the classical beam, and vector modes fall into this category [45-51]. In view of this, in this series, we choose well-established Dirac notation to describe and analyze vector modes and the nonlinear interactions between them. The vectorial nonlinear interaction can thus be regarded as an information-related interaction between (among) qubits (vector light) via an apparatus (nonlinear media). Moreover, in future work, we can readily extend the results here to study quantum nonlinear interactions involving SOC-based hyperentanglement photons.

---

* zhuzhihan@hrbust.edu.cn

In this first paper of the series, we start from the type-II SHG, the simplest nonlinear interaction, to discuss the new phenomena induced by SOC-mediated NPs. The remainder of this paper is organized as follows: We first present a general theoretical framework (Sec. II), based on which we then (Sec. III) analyze the influence of vector pump fields on the spatial structure of the excited NPs as well as the beam profiles of the SHG fields upon propagation. Finally, we experimentally prove the validity and accuracy of the theory (Sec. IV). Both in the theoretical analysis and the experimental demonstration, we consider type-II SHG in a lossless nonlinear optical medium involving collimated and monochromatic input vector beams.

## II. THEORY

### A. Nonlinear polarizations of vectorial type-II SHG

As a typical second-order nonlinear optical process, the NP of the SHG is driven by the quadratic beating of the applied pump field. Specifically, for the type-I SHG driven by a scalar spatial mode $|\psi\rangle = A(r,\varphi,z)|\mathbf{k}(\omega)\rangle$, the corresponding NP can be expressed as $\mathbf{P}^{NL} = 0.5\kappa|\psi\rangle^2 = 0.5\kappa A^2(r,\theta,z)|\mathbf{k}(2\omega)\rangle$, where $\kappa = \epsilon_0 \chi^{(2)}$ is a nonlinear coupling coefficient, and $A(r,\theta,z)$ and $|\mathbf{k}(\omega)\rangle$ denote the spatial complex amplitude and wave-vector term $e^{ik(\omega)z}$ of the mode, respectively. The type-II SHG is a sum-frequency generation between orthogonal, linearly polarized components of the pump field. If we assume that the interaction is phase-matched with respect to the horizontally ($\hat{\mathbf{e}}_H$) and vertically ($\hat{\mathbf{e}}_V$) polarized components of the pump, i.e., the most common case in type-II crystals, the NP created in the crystal by an incident pump with an SoP of $|\hat{\mathbf{e}}_+\rangle = \sqrt{\alpha}|\hat{\mathbf{e}}_H\rangle + e^{i\varphi}\sqrt{1-\alpha}|\hat{\mathbf{e}}_V\rangle$, or explicitly as $|\psi_S\rangle = |\hat{\mathbf{e}}_+,\psi\rangle$, can be expressed as a an SoP-dependent quadratic beating relation:

$$\mathbf{P}^{NL} = \kappa\langle\hat{\mathbf{e}}_H|\psi_S\rangle\langle\hat{\mathbf{e}}_V|\psi_S\rangle$$
$$= \sqrt{\alpha(1-\alpha)}\kappa e^{i\varphi} A^2(r,\theta,z)|\mathbf{k}(2\omega)\rangle, \quad (1)$$

where $\alpha \in [0,1]$ and $e^{i\varphi}$ represent the mode weight and the intramode phase of the SOPs, respectively. From Eq. (1), it is clear, as also shown in Fig. 1 (a), that, first, the spatial profile of the created NP was the quadratic beating of the pump field that led to a shrinking in the waist of the beam of the SHG field (compared with the pump). Second, unlike the type-I SHG, the intensity of the NP, i.e., $I_{NP} = \langle\mathbf{P}^{NL}|\mathbf{P}^{NL}\rangle$, is governed by a variable term $\sqrt{\alpha(1-\alpha)} \leq 0.5$, namely $I_{NP} \propto \alpha(1-\alpha)$. This indicates that the intensity of the type-II SHG completely vanishes for the $\hat{\mathbf{e}}_H$- or $\hat{\mathbf{e}}_V$-polarized pump, and reaches the maximum for circularly ($\hat{\mathbf{e}}_L$ and $\hat{\mathbf{e}}_R$) or diagonally ($\hat{\mathbf{e}}_D$ and $\hat{\mathbf{e}}_A$) polarized pumps. Moreover, the intramodal phase $e^{i\varphi}$ appearing on the right-hand side of Eq. (1) can influence the initial dynamic phase of the generated SHG fields. These well-known facts in scalar nonlinear optics, however, change radically when vector pump fields are involved, as is discussed in more detail below.

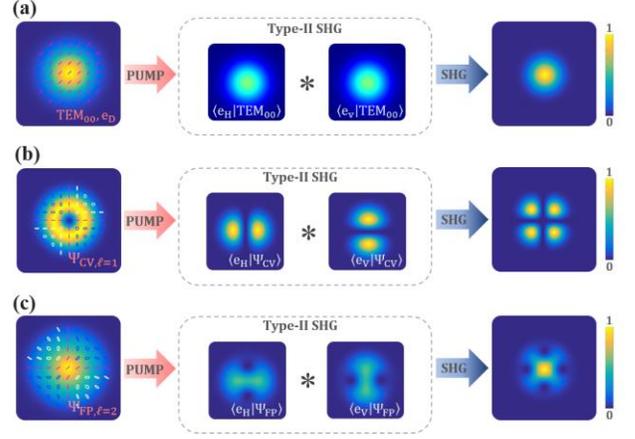

FIG. 1. Schematic representation of the comparison between scalar (a) and vector type-II SHGs, (b) and (c).

The spatially variant SoPs of the vector modes originate from the subwavelength-scale inhomogeneities within the laser fields, i.e., the geometric phase structure. Therefore, the vectorial laser fields can be generally regarded as in a non-separable SOC state with respect to the orthogonal SoPs $\{\hat{\mathbf{e}}_+,\hat{\mathbf{e}}_-\}$ and associated SoP-dependent spatial modes $\{\psi_+,\psi_-\}$, which are respectively given by

$$|\psi_{SOC}\rangle = \sqrt{\beta}|\hat{\mathbf{e}}_+,\psi_+\rangle + e^{i\phi}\sqrt{1-\beta}|\hat{\mathbf{e}}_-,\psi_-\rangle \quad (2)$$

and

$$\begin{cases}|\hat{\mathbf{e}}_+\rangle = \sqrt{\alpha}|\hat{\mathbf{e}}_H\rangle + e^{i\varphi}\sqrt{1-\alpha}|\hat{\mathbf{e}}_V\rangle \\ |\hat{\mathbf{e}}_-\rangle = \sqrt{1-\alpha}|\hat{\mathbf{e}}_H\rangle - e^{i\varphi}\sqrt{\alpha}|\hat{\mathbf{e}}_V\rangle\end{cases}, \quad (3)$$

where $\beta \in [0,1]$ and $e^{i\phi}$ denote the mode weight and associated intramode phase of the SOC state, respectively. For this qubit-like state, we can use the "concurrence" of SOC states to quantify its SOC strength, or rather the degree of non-separability with respect to the orthogonal base of the SOC space [45-51]. This is given by $C_{SOC} = 2\sqrt{\beta(1-\beta)}$, where $C_{SOC} \in [0,1]$ corresponding to the SoP from pure scalar to full vector. According to Eq. (1), we can express the NP of the type-II SHG driven by a vector mode shown in Eq. (2) as

$$\mathbf{P}_{\text{SOC}}^{\text{NL}} = \kappa \langle \hat{\mathbf{e}}_H | \mathbf{\psi}_{\text{SOC}} \rangle \langle \hat{\mathbf{e}}_V | \mathbf{\psi}_{\text{SOC}} \rangle = \kappa e^{i\varphi} |\psi_H\rangle |\psi_V\rangle, \quad (4)$$

with a group of $\hat{\mathbf{e}}_H$- and $\hat{\mathbf{e}}_V$-dependent spatial modes

$$\begin{cases} |\psi_H\rangle = \sqrt{\alpha\beta}|\psi_+\rangle + e^{i\phi}\sqrt{(1-\alpha)(1-\beta)}|\psi_-\rangle \\ |\psi_V\rangle = \sqrt{(1-\alpha)\beta}|\psi_+\rangle - e^{i\phi}\sqrt{\alpha(1-\beta)}|\psi_-\rangle. \end{cases} \quad (5)$$

From Eqs. (3) and (4), we see that the vectorial NP is created by the quadratic beating (or two-wave coupling) of two SoP-dependent spatial modes. Note that $|\psi_H\rangle$ and $|\psi_V\rangle$ are not usually orthogonal to each other, unless $C_{\text{SOC}}=1$. Now, by substituting Eq. (5) into Eq. (4), we can reformulate the NP as

$$\begin{aligned} \mathbf{P}_{\text{SOC}}^{\text{NL}} &= \kappa e^{i\varphi}\left(a_1|\psi_+\rangle^2 + a_2|\psi_-\rangle^2 + a_3|\psi_+\rangle|\psi_-\rangle\right) \\ &= \kappa e^{i\varphi}\left(a_1|\psi_1\rangle + a_2|\psi_2\rangle + a_3|\psi_3\rangle\right) \quad , \end{aligned} \quad (6)$$

where $|\psi_{1,2,3}\rangle$ denote various spatial-mode components of the created NP, and their corresponding complex probability amplitudes $a_{1,2,3}$ are given by

$$\begin{cases} a_1 = \sqrt{\alpha(1-\alpha)}\beta \\ a_2 = e^{i(2\phi+\pi)}\sqrt{\alpha(1-\alpha)}(1-\beta) \\ a_3 = e^{i\phi}(1-2\alpha)\sqrt{\beta(1-\beta)} \end{cases}. \quad (7)$$

Note that the components $|\psi_{1,2,3}\rangle$ shown in Eq. (6) are spatial modes constructed by two-wave couplings $|\psi_+\rangle^2$, $|\psi_-\rangle^2$, and $|\psi_+\rangle|\psi_-\rangle$, respectively. The degree of coupling depends on their overlap in the transverse plane. For the third component, owing to the possible difference between $\langle\psi_+|\psi_+\rangle$ and $\langle\psi_-|\psi_-\rangle$ in plane $\{r,\varphi\}$, we have a relation $\iint\langle\psi_3|\psi_3\rangle dr d\theta \in [0,1]$. We now introduce parameters $C_S = 2\sqrt{\alpha(1-\alpha)}$ and $P_S = \sqrt{1-C_S^2}$, and can then reformulate $a_{1,2,3}$ as $a_1 = 0.5C_S\beta$, $a_2 = 0.5e^{i(2\phi+\pi)}C_S(1-\beta)$, and $a_3 = 0.5e^{i\phi}P_S^2 C_{\text{SOC}}$, respectively. The two parameters quantify the "concurrence" and "polarizability" of the SoPs with respect to the base $\{\hat{\mathbf{e}}_H, \hat{\mathbf{e}}_V\}$, and the quadratic constraint $C_S^2 + P_S^2 = 1$ relates them to the coherence of polarization of the light fields [52,53]. Remarkably, in the following, we see that $C_S$ and $C_{\text{SOC}}$ have an equivalent effect on $I_{\text{NP}}$.

Equations (4)–(7) provide a general description of the type-II SHG driven by a vector mode, and Eqs. (6) and (7) represent the selection rule for the spatial structure of the NP created by the vector modes. From the above equations, we find that the interaction is governed fully by the SOC state of incident pump, or rather by the coupling of SoP-dependent spatial modes $|\psi_H\rangle|\psi_V\rangle$, as two specific examples shown in Figs. 1 (b) and (c). In this new realm, the SOC state of the applied pump plays a vital role in vectorial interaction—i.e., it directly determines both the spatial structure and the intensity of the NP created in the crystal. For the selection rule shown in Eqs. (6) and (7), it is important to note that the conversion efficiency of a nonlinear process is proportional to the power density of the driven pump. This indicates that the three spatial mode components, owing to possible difference in power density (related to spatial size), may have different SHG efficiencies, such as NPs created by the full Poincaré (FP) modes discussed in Sec. III B. That is, the spatial structure of the NPs (or the generated SHG waves) is determined not only by $a_{1,2,3}$ but also by the power density of $|\psi_{1,2,3}\rangle$.

We now discuss the intensity of the NPs shown in Eq. (6), which is proportional to the power of the SHG wave obtained. In this section, we consider only the case where two SoP-dependent spatial modes $|\psi_+\rangle$ and $|\psi_-\rangle$ have the same spatial profile, i.e., $\langle\psi_+|\psi_+\rangle = \langle\psi_-|\psi_-\rangle$, such as the most common cylindrical vector (CV) modes. Under this restriction, $I_{\text{NP}} \propto a_1^2 + a_2^2 + a_3^2$ indicating that $I_{\text{NP}}$ is completely determined by the complex probability amplitudes. To illustrate this common case, Fig. 2 shows the normalized $a_{1,2,3}$ and its associated $I_{\text{NP}}$ as functions of $\alpha$ and $\beta$, respectively. We see that, first, the content of the spatial mode and $I_{\text{NP}}$ are dominated by both $C_{\text{SOC}}$ and $C_S$, and, second, that unlike in the scalar case, $I_{\text{NP}}(\alpha,\beta)$ is always nonzero in the vector case as $\beta \neq 0,1$. This is because any vector mode ($C_{\text{SOC}} \neq 0$) can be expressed with respect to the base $\{|\hat{\mathbf{e}}_H,\psi_H\rangle,|\hat{\mathbf{e}}_V,\psi_V\rangle\}$, and has a relation $|\psi_H\rangle|\psi_V\rangle \neq 0$. Moreover, according to Fig. 2, when the SOC strength of the pump tends toward zero, i.e., $C_{\text{SOC}} = 0$ corresponding to $\beta = 0$ or 1, the changes in $a_{1,2,3}$ and $I_{\text{NP}}$ with $\alpha$ obey the rule for the pure scalar case shown in Eq. (1), i.e., $a_1$ or $a_2 \equiv 0$, $a_3 \equiv 0$, and $I_{\text{NP}} \propto \alpha(1-\alpha)$. These results show that Eqs. (5) and (6) provide a unified description of the type-II SHG that is compatible with both the scalar and vector cases.

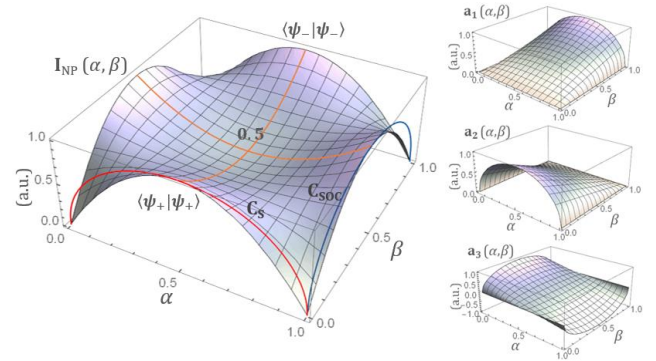

FIG. 2. Normalized complex probability amplitudes of the three components of the spatial mode (right) and the corresponding NP intensity (left) as functions of parameters $\alpha$ and $\beta$, respectively. In the diagram of $I_{\text{NP}}(\alpha,\beta)$, the red and blue curves are functions $C_S(\alpha)$ and $C_{\text{SOC}}(\beta)$, respectively.

### B. Beam profiles of generated second harmonic fields upon diffraction propagation

In the SHG process, the NP discussed above directly drives the dipole moment in the media to generate a new traveling wave with a frequency of $2\omega$, i.e., the SHG field. Therefore, the SHG field during nonlinear propagation can be calculated through an inhomogeneous wave equation with the corresponding NP as driven terms:

$$\nabla^2 \mathbf{E} - \frac{n^2}{c^2}\frac{\partial^2 \mathbf{E}}{\partial t^2} = \frac{1}{\epsilon_0 c^2}\frac{\partial^2 \mathbf{P}_{\text{SOC}}^{\text{NL}}}{\partial t^2}. \quad (8)$$

From Eq. (8), we can obtain a well-known coupled-wave equation set that can predict the amplitude of the pump and SHG fields upon nonlinear propagation [3]. We are interested here in the spatial structures of second harmonic fields generated from the generation plane $z_0$ (in the crystal) to the far field. The SOC structure of the residual pump field in case of pump depletion is discussed in our subsequent papers. For this, we find that the driven term on the right-hand side of Eq. (8) acts as the wave source of the SHG fields generated in the crystal [3]. Therefore, we can use the diffraction-integral (Green function) method with the corresponding NP wavefunction as pupil function to conveniently calculate the diffractive property (or beam profile) of the SHG field upon propagation. The spatial modes of vector light considered here are paraxial laser beams. In view of this, we choose the Collins propagator to derive the spatial wavefunction of the SHG fields upon propagation in $z$, which is given by [54]:

$$\mathbf{E}_{\text{SHG}}(r,\varphi,z) = \frac{i}{\lambda z}\exp[-ikz]\int r_0 dr_0 \int d\varphi_0 \mathbf{E}_{\text{pupil}}(r_0,\varphi_0,z_0) \\ \times \exp\left\{-\frac{ik}{2z}[r_0^2 - 2rr_0\cos(\varphi-\varphi_0)+r^2]\right\}, \quad (9)$$

where $\lambda$ denotes the wavelength of the SHG field, $\mathbf{E}_{\text{pupil}}(r_0,\varphi_0,z_0) = |\psi_H\rangle|\psi_V\rangle$ is the pupil function of the diffraction integral that is equal to the wavefunction of the SHG field at $z_0$. Note that in this work, we assume that nonlinear interaction is limited within a short interval of the Rayleigh distance ($z_R$). Nonlinear interactions with focused vector beams, i.e., the interactions occurring within several orders of $z_R$, are the subject of a separate paper in this series.

## III. SIMULATION ANALYSIS

In this section, we employ the theoretical toolkit provided in the previous section to explore how the SOC of two categories of vector modes, i.e., (i) cylindrical vector (CV) modes and (ii) full Poincaré (FP) modes, dictate the type-II SHG process. We focus on the spatial structure (spatial mode spectrum) of the created NP as a function of $\alpha$ and $\beta$ (or $C_{\text{SOC}}$ and $C_{\text{S}}$) as manifested in two experimental observables: (1) the intensity of the created NP that is proportional to the output power of the obtained SHG, and (2) the corresponding beam profiles of the generated second harmonic fields upon propagation.

For both derivation and simulation, we chose the Laguerre–Gauss (LG) mode as spatial mode carrying the OAM (see Eq. (A1) in Appendix A). In the following, we represent LG modes as $|LG_{\pm\ell}\rangle = A_{|\ell|}(r,\varphi,z)|\mathbf{k}(\omega),\pm\ell\rangle$, where $A_{|\ell|}(r,\varphi,z)$ (we use $A_{|\ell|}$ for short in the following) is the donut-like spatial amplitude of the LG modes with topological charges $\pm\ell$, and $|\ell\rangle$ denotes the vortex wavefront $e^{i\ell\varphi}$ (i.e., OAM-carrying term). Note that in a given propagation plane, two complementary LG modes $|LG_{+\ell}\rangle$ and $|LG_{-\ell}\rangle$ have the same spatial amplitude $A_{|\ell|}$.

In our experiment, the initial SOC states were prepared as $\sqrt{\beta}|\hat{\mathbf{e}}_H,\psi_H\rangle + e^{i\phi}\sqrt{1-\beta}|\hat{\mathbf{e}}_V,\psi_V\rangle$ (see the experimental setup in Fig. 7), and the SoP base was further manipulated to the desired base $\{\hat{\mathbf{e}}_+,\hat{\mathbf{e}}_-\}$ using a half-wave plate (HWP) and a quarter-wave plate (QWP). Therefore, to conveniently compare the results of the simulation with those of the experiment, we assumed in the former that the SoP base $\{\hat{\mathbf{e}}_+,\hat{\mathbf{e}}_-\}$ was converted from $\{\hat{\mathbf{e}}_H,\hat{\mathbf{e}}_V\}$ via a HWP or a QWP. That is, the weight coefficient $\alpha$ was regarded as a function of the angle of the HWP $\delta_{1/2}$ or QWP $\delta_{1/4}$ (see details in Appendix B). Figure 3(a) shows the function of $\alpha$ with respect to $\delta_{1/2}$ and $\delta_{1/4}$, where we see that a change in $\delta_{1/2}(0° \to 45°)$ corresponds to $\alpha(1 \to 0)$ and that in $\delta_{1/4}(0° \to 45° \to 90°)$ corresponds to $\alpha(1 \to 0.5 \to 1)$.

### A. Vectorial type-II SHG driven by CV-mode pump

The so-called CV modes refer to the most common category of vector spatial modes that feature donut-like intensity profiles and spatially variant SoPs, and are therefore also known as vector vortices [26-28]. The SOC state of the CV modes can be described as a non-separable superposition of two opposite LG modes with mutually orthogonal SoPs given by

$$|\psi_{\text{CV}}\rangle = \sqrt{\beta}|\hat{\mathbf{e}}_+,LG_{+\ell}\rangle + e^{i\phi}\sqrt{1-\beta}|\hat{\mathbf{e}}_-,LG_{-\ell}\rangle \\ = A_{|\ell|}|\mathbf{k}(\omega)\rangle\left(\sqrt{\beta}|\hat{\mathbf{e}}_+,+\ell\rangle + e^{i\phi}\sqrt{1-\beta}|\hat{\mathbf{e}}_-,-\ell\rangle\right). \quad (10)$$

Equation (10) indicates that the CV modes possess a donut-like spatial amplitude $A_{|\ell|}$ and an associated CV-type vector wavefront $\sqrt{\beta}|\hat{\mathbf{e}}_+,+\ell\rangle + e^{i\phi}\sqrt{1-\beta}|\hat{\mathbf{e}}_-,-\ell\rangle$. The substitution of Eq. (10) into Eq. (6) provides an expression to describe the NP created by CV-mode pumps:

$$\mathbf{P}_{\text{CV}}^{\text{NL}} = \kappa\langle\hat{\mathbf{e}}_H|\psi_{\text{CV}}\rangle\langle\hat{\mathbf{e}}_V|\psi_{\text{CV}}\rangle \\ = \kappa e^{i\varphi}\left(a_1|LG_{+\ell}\rangle^2 + a_2|LG_{-\ell}\rangle^2 + a_3|LG_{+\ell}\rangle|LG_{-\ell}\rangle\right) \\ = \kappa e^{i\varphi}A_{|\ell|}^2|\mathbf{k}(2\omega)\rangle\left(a_1|+2\ell\rangle + a_2|-2\ell\rangle + a_3|0\rangle\right). \quad (11)$$

We see that for a given topological charge, the created NP has a definite radial amplitude structure $A_{|\ell|}^2$ but a variable azimuthal phase structure $a_1|+2\ell\rangle + a_2|-2\ell\rangle + a_3|0\rangle$. In other words, the three components of the NP, i.e., $|2\ell\rangle$, $|-2\ell\rangle$, and $|0\rangle$, have the same intensity profile $A_{|\ell|}^2$. As a consequence, the spatial structure of the generated SHG field depends only on the complex probability amplitudes $a_{1,2,3}$. That is, as mentioned in Sec. II, the SHG pumped by the CV modes is controlled fully by the parameters of the pump field $\alpha$ and $\beta$ (or $C_{\text{SOC}}$ and $C_{\text{S}}$).

To demonstrate the prediction shown in Fig. 2, we start from the special case of $\beta = 0.5$ corresponding to $C_{SOC} = 1$ when the pump fields are full vectorial [47-49]. Moreover, for simplicity and without loss of generality, we set the intramodal phase $\phi$ to zero. Based on the relations shown in Fig. 3(a), first, Figs. 3 (d) and (e) show the simulated pump fields featuring a CV-type SOC structure versus $\delta_{1/2}(0° \rightarrow 45°)$ and $\delta_{1/4}(0° \rightarrow 90°)$, respectively (first row). The spatial profiles of their corresponding $|\psi_H\rangle$ and $|\psi_V\rangle$ are shown in the middle row, while the spatial profiles of the created NPs are shown in the bottom row. From these simulations, we see that the spatial structure of $|\psi_H\rangle$ and $|\psi_V\rangle$ are governed fully by the $C_S$ of the applied pump, which leads further to a SoP-controlled NP creation. On the basis of these simulated profiles, for a more intuitive comparison, Figures 3 (b) and (c) show the corresponding $I_{NP}$ as a function of $\delta_{1/2}$ and $\delta_{1/4}$ with different $\ell$, respectively. The results illustrate that as predicted in Sec. II, first, $I_{NP}$ remains nonzero as $C_{SOC} \neq 0$, and, second, due to the increase in the beam size of the CV modes upon $\ell$, leading to a decrease in the power density of the pump, $I_{NP}$ decreases with the increase of topological charges.

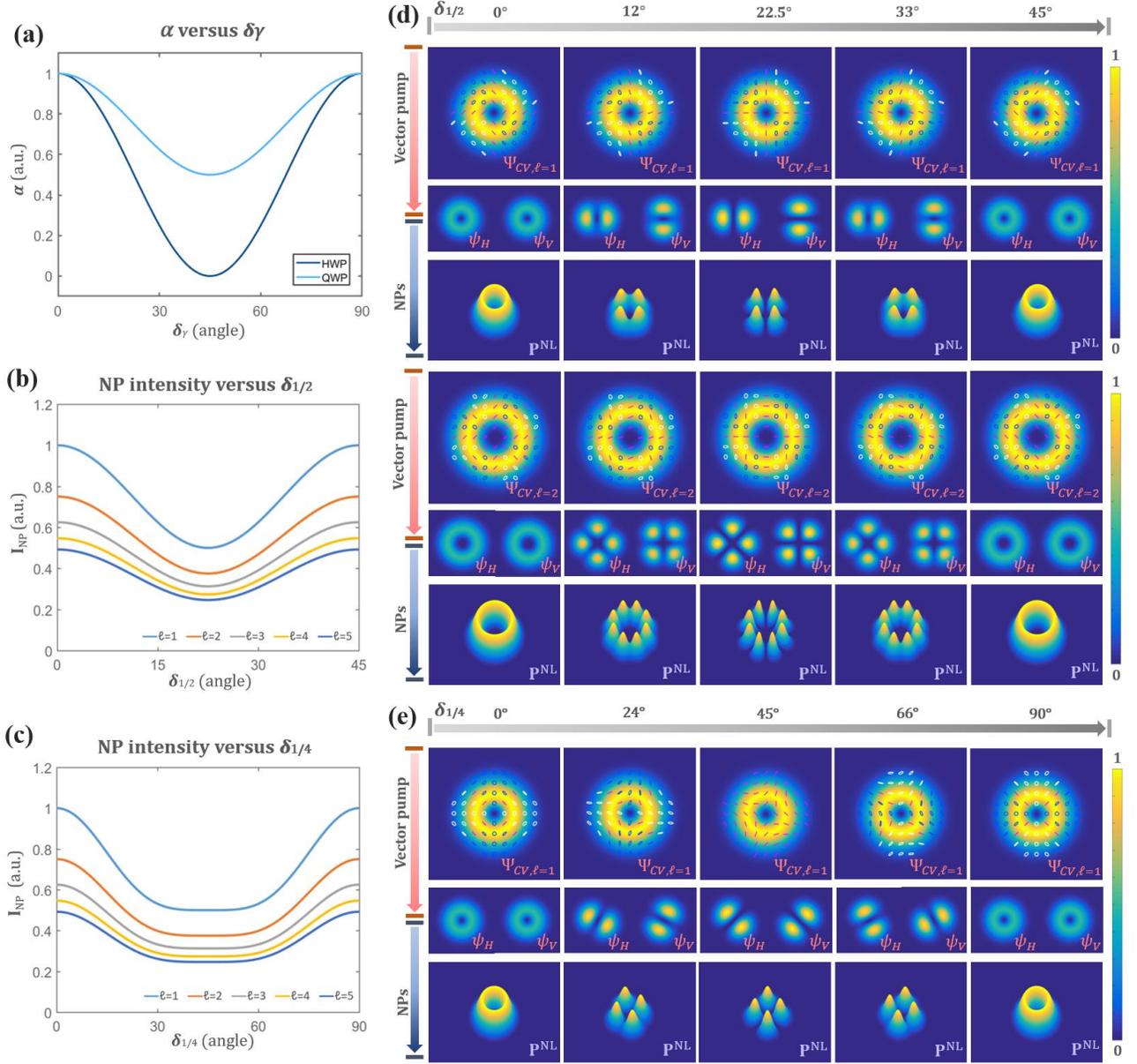

FIG. 3. (a) $\alpha$ as a function of $\delta_{1/2}$ and $\delta_{1/4}$, respectively. (b) and (c) are the simulated intensities of the NPs as function of $\delta_{1/2}$ and $\delta_{1/4}$ for $\ell = 1 \sim 5$, respectively. (d) and (e) depict the simulated SOC structures of the CV-mode pumps for different angles $\delta_{1/2}$ and $\delta_{1/4}$, their corresponding SoP-dependent orthogonal spatial mode pairs, and spatial profiles of NPs driven by them. The blue (white) circles and red lines on the simulated SOC structures correspond to left (right) circular and linear SoPs, respectively.

The results shown in Fig. 3 indicate that when the value of $C_{SOC}$ of the pump is constant (i.e., $\beta = 0.5$), the spatial structure of the created NP is fully controlled by the value of $C_S$ (or $\alpha$) of the applied pump. In addition, in the simulated profiles shown in Figs. 3 (b) and (c), of particular note are the two specific cases (i) $\delta_{1/2} = 0°, 45°$ or $\delta_{1/4} = 0°, 90°$ corresponding to $\alpha = 0, 1$, and (ii) $|HG_{01}\rangle = \sqrt{1/2}(|+\ell\rangle - |-\ell\rangle)$ and $\delta_{1/4} = 45°$ corresponding to $\alpha = 0.5$. For case (i), the SHG can be regarded as being pumped by $\sqrt{1/2}A_{|\ell|}|\mathbf{k}(\omega)\rangle(|\hat{\mathbf{e}}_H,+\ell\rangle+|\hat{\mathbf{e}}_V,-\ell\rangle)$, and, according to Eq. (6), the created NP can be expressed as

$$\mathbf{P}_{CV}^{NL} = 0.5\kappa A_{|\ell|}^2 |\mathbf{k}(2\omega)\rangle |0\rangle, \tag{12}$$

corresponding to $a_1 = a_2 = 0$ and $a_3 = 0.5$. Note that in this case, $I_{NP}$ reaches its maximum value owing to the perfect overlap between $|\psi_H\rangle$ and $|\psi_V\rangle$. For case (ii), the driven pumps can be regarded as $\sqrt{1/2}A_{|\ell|}(|\hat{\mathbf{e}}_H,HG_{10}\rangle+|\hat{\mathbf{e}}_V,HG_{01}\rangle)$ and $\sqrt{1/2}A_{|\ell|}(|\hat{\mathbf{e}}_L,+\ell\rangle+i|\hat{\mathbf{e}}_R,-\ell\rangle)$, where $|HG_{10,01}\rangle = \sqrt{1/2}(|+\ell\rangle \pm |-\ell\rangle)$ denotes the Hermite–Gauss phases. Similarly, according to Eq. (6), we can express the created NPs as

$$\mathbf{P}_{CV}^{NL} = 1/4\kappa A_{|\ell|}^2 |\mathbf{k}(2\omega)\rangle(|+2\ell\rangle - |-2\ell\rangle) \tag{13}$$

and

$$\mathbf{P}_{CV}^{NL} = -1/4 i\kappa A_{|\ell|}^2 |\mathbf{k}(2\omega)\rangle(|+2\ell\rangle + |-2\ell\rangle), \tag{14}$$

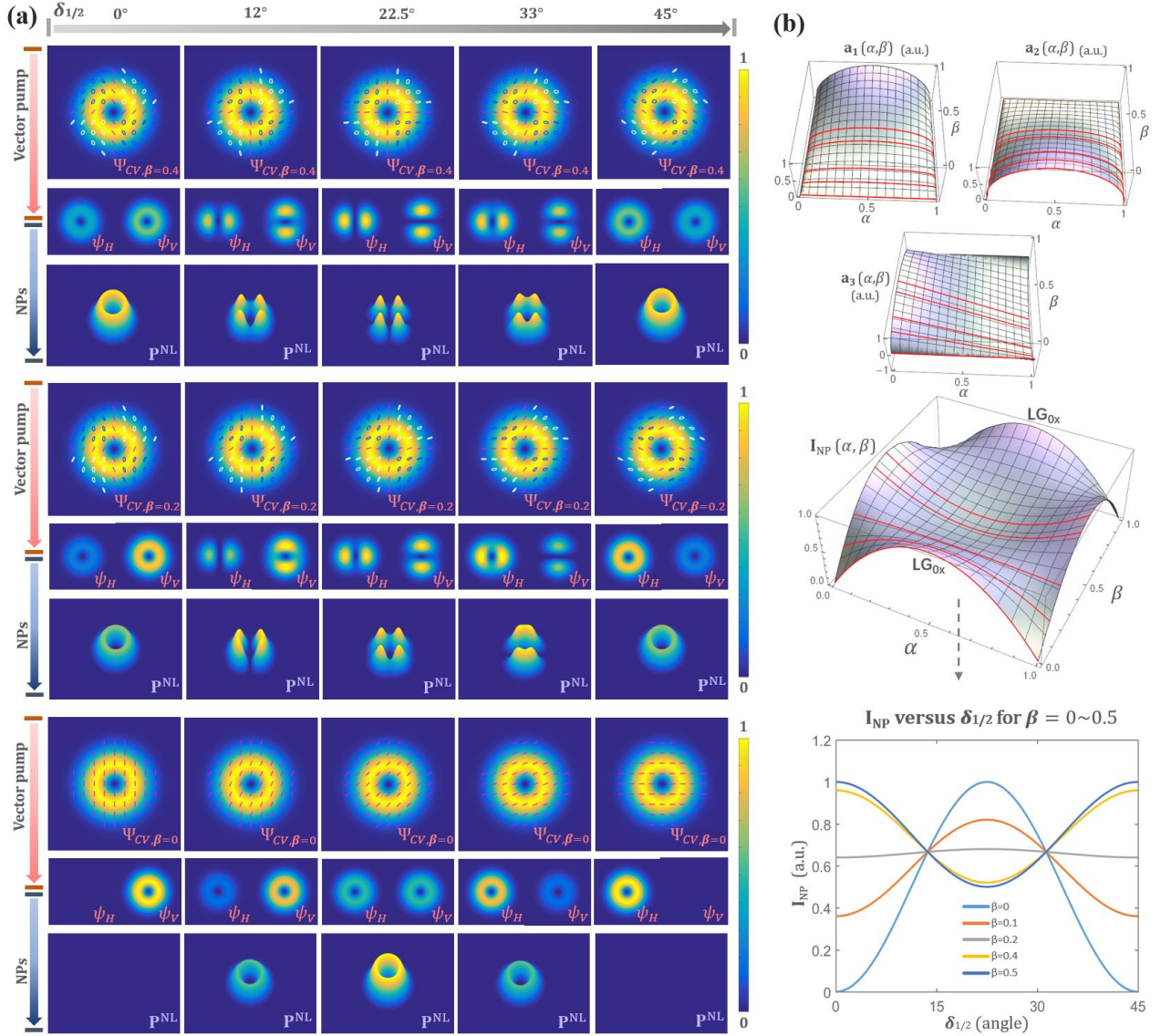

FIG. 4. (a) Simulated SOC structures of CV-mode pumps for different angles $\delta_{1/2}$, their corresponding SoP-dependent orthogonal spatial mode pairs, and spatial profiles of the NPs driven by them. We set $\ell = 1$, and the three rows correspond to $\beta = 0.4$, $0.2$ and $0$. (b) The first two rows show $a_{1,2,3}$ and the corresponding $I_{NP}$ as functions of $\alpha$ and $\beta$, respectively; and the third row shows $I_{NP}$ as a function of $\delta_{1/2}$ for $\beta = 0 \sim 0.5$.

respectively. Note that $I_{NP}$ takes its minimum value, i.e., half the maximum value obtained in case (i), as the overlaps between $|\psi_H\rangle$ and $|\psi_V\rangle$ are minimal.

We further consider the common case where the driven pump's $C_{SOC} < 1$, i.e., $\beta \neq 0.5$. At this time, the spatial structures of $|\psi_H\rangle$ and $|\psi_V\rangle$ as well as the NP created by them are codetermined by $C_{SOC}$ and $C_S$ of the applied pump. In particular, the specific cases $\delta_{1/2} = 0°, 22.5°$, shown in Eqs. (12) and (13), can now be rewritten as

$$\mathbf{P}_{CV}^{NL} = \sqrt{\beta(1-\beta)}\kappa A_{|\ell|}^2 |\mathbf{k}(2\omega)\rangle |0\rangle, \qquad (15)$$

and

$$\mathbf{P}_{CV}^{NL} = \frac{1}{2}\kappa A_{|\ell|}^2 |\mathbf{k}(2\omega)\rangle \left[ \beta |+2\ell\rangle - (1-\beta)|-2\ell\rangle \right], \qquad (16)$$

respectively. Figure 4(a) shows the simulated SOC structures of the partially vectorial (or entangled) CV-mode pump versus $\delta_{1/2}(0° \to 45°)$, where $\beta = 0.4, 0.2, 0$ (upper row). Their corresponding values of $|\psi_H\rangle$ and $|\psi_V\rangle$ are shown in the middle row. The spatial profiles of the created NPs are shown in the bottom row. Moreover, the plots shown in Fig. 4(b) present $a_{1,2,3}$ and the corresponding $I_{NP}$ of the created NPs as functions of $\alpha$, $\beta$, and $\delta_{1/2}$ for $\beta = 0 \sim 0.5$. Fig. 4 shows that the vectorial feature manifested in $|\psi_H\rangle$ and $|\psi_V\rangle$ as well as the NP excited by them wears off with a decrease in the strength of the SOC ($C_{SOC}$) of the pump field. As a result, the change in $a_{1,2,3}$ and the corresponding $I_{NP}$ with $\alpha$ gradually evolves into the case of a scalar SHG obeying Eq. (1).

On the basis of the NPs discussed above, we can readily predict the beam profiles of the corresponding SHG fields originating from them using the diffraction integral shown in Eq. (9). Note that in Figs. 3 and 4, the initial intensity and phase structures of the NPs created at points $\delta_{1/2} = 22.5°$ and $\delta_{1/4} = 45°$ match each other and feature a petal-like profile with orthogonally composed, twisted phase structures. In contrast, they do not match at points $\delta_{1/2} = 0°, 45°$ and $\delta_{1/4} = 0°, 90°$, i.e., they feature a donut profile without twisted phase structures. This has a profound influence on the beam profiles of the SHG fields during diffraction propagation. In light of this, we focus on the far-field SHG beam profiles originating from NPs of these two categories.

We first consider the SHG pumped by fully entangled CV modes, and Fig. 5(a) presents the simulated far-field ($z_R = 3$) SHG beam profiles originating from $\mathbf{P}_{CV}^{NL}(\delta_{1/2} = 0°, 45°)$ and $\mathbf{P}_{CV}^{NL}(\delta_{1/2} = 22.5°)$ with $\ell = 1, 3$, and $5$, respectively. For the first category, the beam profiles gradually evolve into $TEM_{00}$-like distributions with a faint outer-ring texture in the far field, and the complexity of the texture increases with the topological charge carried by the pump field. These results coincide with the experimental proof shown in Sec. IV and, to some extent, are also consistent with Pereira et al.'s prediction in Ref. [55]. For the second category, by contrast, the SHG beam profiles remain constant upon propagation because they originate from the NPs. We can interpret these results as follows: First, both the intensity and the phase structures of the NPs were tailored by the SOC state of the pump fields, i.e., the beam profile of the SHG light at the nonlinear interaction plane $z_0$ (wave source). Second, if the initial intensity and phase structures of the SHG light match each other, the beam profile remains constant upon propagation like that of paraxial eigenmodes. Otherwise, a drastic evolution in the beam profile occurs during diffractions. We further consider the influence of $C_{SOC}$ of the pump on the SHG beam profile. Figure 5(b) presents the simulated far-field ($z_R = 3$) SHG beam profiles originating from $\mathbf{P}_{CV}^{NL}(\delta_{1/2} = 0°, 45°)$ and $\mathbf{P}_{CV}^{NL}(\delta_{1/2} = 22.5°)$ (with $\ell = 1$) for $\beta = 0, 0.2$, and $0.4$. We find the same behavior again, i.e., the beam profile of the first category is unstable upon propagation while the other remains constant. However, the difference is that with decreasing $C_{SOC}$, the beam profiles both degenerate into that of scalar type-II SHG, i.e., they gradually vanish or transform into a donut shape.

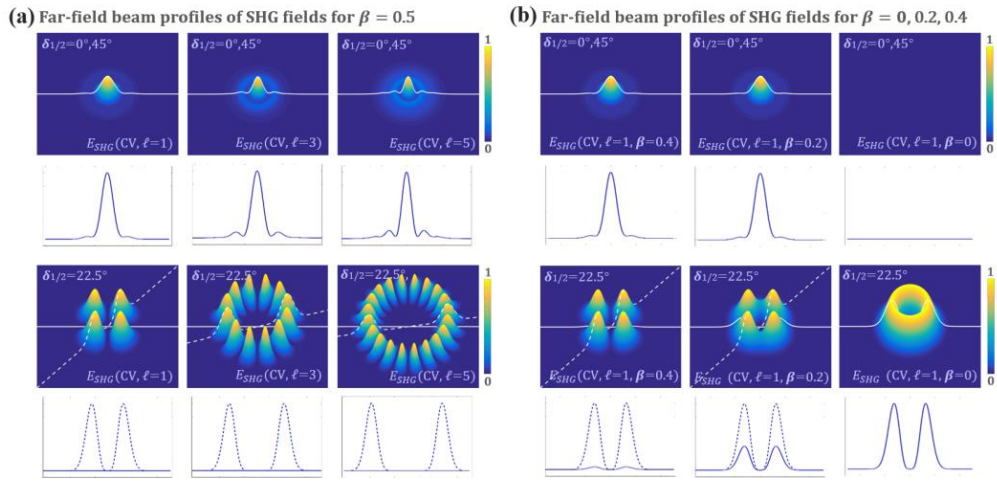

FIG. 5. Simulated far-field SHG beam profiles originating from $\mathbf{P}_{CV}^{NL}(\delta_{1/2} = 0°, 45°)$ and $\mathbf{P}_{CV}^{NL}(\delta_{1/2} = 22.5°)$, where (a) corresponds to $\beta = 0.5$ and $\ell = 1, 3$, and $5$, and (b) corresponds to $\ell = 1$ and $\beta = 0.4 \sim 0$.

## B. Vectorial type-II SHG driven by FP-mode pump

We now analyze the SHG driven by the so-called FP modes whose local SoPs in the transverse plane can cover at least one surface of a Poincaré sphere [26-29]. This unique polarization structures arises by considering the spatial profiles of OAM-carrying modes with different values of $|\ell|$. We consider the most common case where one spatial mode is the TEM$_{00}$ mode, i.e., $\ell = 0$, that can be expressed as:

$$|\psi_{FP}\rangle = \sqrt{\beta}|\hat{e}_+, LG_0\rangle + e^{i\phi}\sqrt{1-\beta}|\hat{e}_-, LG_\ell\rangle$$
$$= \left(\sqrt{\beta}A_0|\hat{e}_+, 0\rangle + e^{i\phi}\sqrt{1-\beta}A_{|\ell|}|\hat{e}_-, \ell\rangle\right)|k(\omega)\rangle, \quad (17)$$

where $A_0$ denotes the intensity profile of the TEM$_{00}$ mode. Equation (17) indicates that the FP mode has an intensity profile $\sqrt{\beta}A_0 + \sqrt{1-\beta}A_{|\ell|}$ with an FP-type vector wavefront $\sqrt{\beta}|\hat{e}_+, 0\rangle + e^{i\phi}\sqrt{1-\beta}|\hat{e}_-, \ell\rangle$. Similar to our previous analysis, by substituting Eq. (17) into Eq. (6), we can obtain the NP created by the FP-mode pumps, given by

$$\mathbf{P}_{FP}^{NL} = \kappa\langle\hat{e}_H|\psi_{FP}\rangle\langle\hat{e}_V|\psi_{FP}\rangle$$
$$= \kappa e^{i\phi}\left(a_1|LG_0\rangle^2 + a_2|LG_\ell\rangle^2 + a_3|LG_0\rangle|LG_\ell\rangle\right)$$
$$= \kappa e^{i\phi}|k(2\omega)\rangle\left(a_1 A_0^2|0\rangle + a_2 A_{|\ell|}^2|2\ell\rangle + a_3 A_0 A_{|\ell|}|\ell\rangle\right). \quad (18)$$

Note that the spatial profiles of $|LG_0\rangle$ and $|LG_\ell\rangle$ are no longer the same, and this leads to a difference in power density between $A_0^2$, $A_{|\ell|}^2$ and $A_0 A_{|\ell|}$. In this special case, as mentioned in Sec. II, the three components of the spatial mode, i.e., $|0\rangle$, $|2\ell\rangle$, and $|\ell\rangle$, have different SHG efficiencies. As a consequence, the spatial structure and intensity of the created NP is determined by $a_{1,2,3}$ and the power densities of $A_0^2$, $A_{|\ell|}^2$ and $A_0 A_{|\ell|}$.

Figure 6 (a) shows the simulated $I_{NP}(\alpha, \beta)$ for $\ell = 1, 2$ as well as a comparison of $I_{NP}(\delta_{1/2})$ for $\ell = 1, 2$ as $\beta = 0.5$. The result shows that similarly to the case of the previous CV mode, $I_{NP}$ created by the FP-mode pump remained nonzero as $C_{SOC} \neq 0$ for arbitrary $\delta_{1/2}$. But the difference is that owing to the wide difference in the power densities of LG modes with different values of $|\ell|$, the impacts of $\beta = 0$ and $\beta = 1$ on $I_{NP}$ are no longer identical. Thus, as shown in Fig. 6 (a), compared with the case where $\ell = 1$, the curved shape of $I_{NP}(\alpha)$ for $\ell = 2$ as $\beta = 0.5$ is reversed. Figure 6 (b) shows the simulated SOC structures of the applied FP-mode pump fields for $\delta_{1/2}(0° \to 45°)$ (upper row), where we have set $\beta = 0.5$ for simplicity, the spatial profiles of $|\psi_H\rangle$ and $|\psi_V\rangle$ (middle row), and the NPs created by them (bottom row). According to Eq. (18), the initial intensities and phase structures of the created NPs match with each other, indicating that their beam profiles remain constant upon propagation. In addition, except for the case where $\delta_{1/2} = 0°, 45°$, the generated SHG fields both carry multi topological charges. Therefore, due to $|\ell|$-dependent Gouy phase accumulation, the SHG beam profiles experience a $90°$ rotation from $z_0$ to the far field [35].

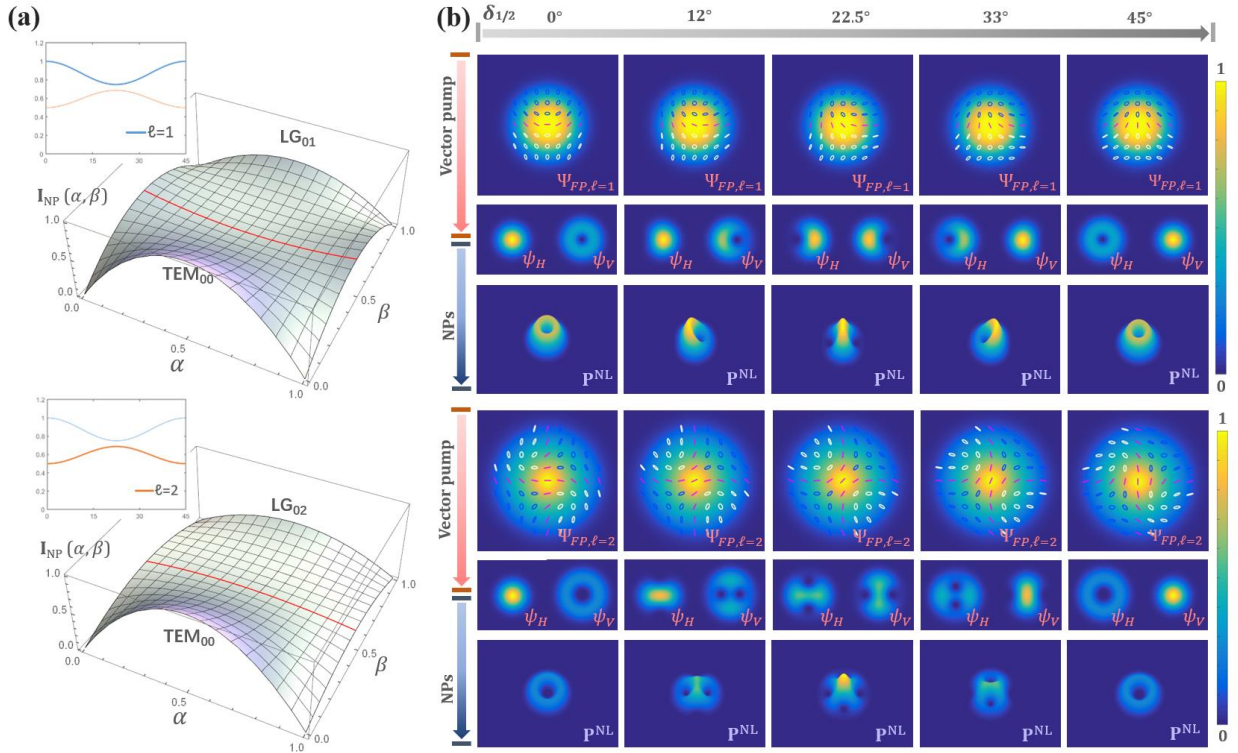

FIG. 6. (a) Simulated NP intensities as a function of $\alpha$ and $\beta$, for $\ell = 1$ and $2$. (b) Simulated SOC structures of FP-mode pumps for different angles $\delta_{1/2}$, their corresponding SoP-dependent orthogonal spatial mode pairs, and spatial profiles of NPs driven by them.

## IV. EXPERIMENTAL DEMONSTRATION

To verify the validity and accuracy of the theoretical method and analysis, in this section, we experimentally verify the above predictions. We focus on the two observables that can prove features of the corresponding NPs predicted in Sec. III: (i) $I_{NP}$ created by the corresponding vector pump that can be measured by comparing the output power of obtained SHG, and (ii) beam profiles of the generated second harmonic fields from the generation plane to the far field.

Figure 7 shows a schematic representation of our experimental setup. A horizontally polarized $TEM_{00}$ mode beam at 800 nm (Toptica TA pro) was converted to the desired spatial mode by using a spatial light modulator (SLM, HOLOEYE PLUTO-2-NIR-80) in combination with a polarizing beam splitter (PBS) and an HWP. For CV-mode generation, we used the so-called "complex amplitude modulation" method to generate pure LG modes from a $TEM_{00}$-mode illumination laser [56]. The generated LG mode was then injected into a polarization (two-arm) Sagnac interferometer containing a Dove prism in one of the paths to convert it into the desired CV mode. An HWP at the input port of the interferometer was used to control the ratio of laser power of the two arms to manipulate the parameter $\beta$ of the generated CV mode. Moreover, at the output port of the Sagnac interferometer, an HWP in combination with a QWP was used to manipulate the parameter $\alpha$ of the CV mode on demand.

The method for FP-mode generation is shown in the upper-left inset of Fig. 7. By utilizing the $\hat{e}_H$-only modulation property of liquid crystal-based SLM, an incident $TEM_{00}$ beam with diagonal polarization was conveniently converted into the FP mode. Because in FP mode preparation, the phase-only modulation was used to load the twisted phase into the incident light, the converted OAM-carrying mode was the hyper-geometric Gaussian mode containing a propagation-varying radial structure [57, 58]. This generated a larger difference between the observed and the simulated results, especially for the case in which the pump light carried larger topological charges, as shown in the experimental results in Fig. 9. In the vectorial SHG process, the generated vector pump was focused into a 5-mm-long type-II PPKTP using a 200 mm focal length lens to drive the SHG. A dichroic mirror (DM) was then used to filter the generated 400 nm SHG fields. A 4*f*-imaging system was employed to image the beam profiles of the SHG fields from the NP (generation) plane to the far field, and the corresponding beam profiles and beam powers were recorded by a CCD (Laser Beam Profiler) mounted on a translation stage.

In the experiment, the beam profile of the SHG field driven by the CV-mode pumps was first observed. We first considered the full vectorial pump (i.e., $\beta = 0.5$). The observed beam profiles (eight-bit false-color greyscale) for different angles $\delta_{l/2}$ upon propagation, ranging from the generation plane $z_0$ (i.e., the intensity profile of the NP) and the intermediate stage to the far field ($z_R = 3$), are shown in Figs. 8 (a1), (a2), and (a3) corresponding to $\ell = 1, 2$ and $3$, respectively. For the SHG driven by a partial vector pump, we considered the case where $\beta = 0.2$, and Fig. 8 (b1) and (b2) show the corresponding experimental results. For convenience of comparison, we also present the simulated beam profiles, shown in Figs. 8(a1*)–(b2*), with the same false-color greyscale that correspond to the experimental observations. The experimental observations exactly match the theoretical predictions shown in the right column of Fig. 8. For clarity, Figs. 8 (c) and (d) present the theoretical spatial mode spectrum of the generated SHG fields as a function of $\delta_{l/2}$. Note that the spatial mode spectra of the SHG fields pumped by CV modes carrying different $\ell$ are identical as mentioned in Secs. II and III (A).

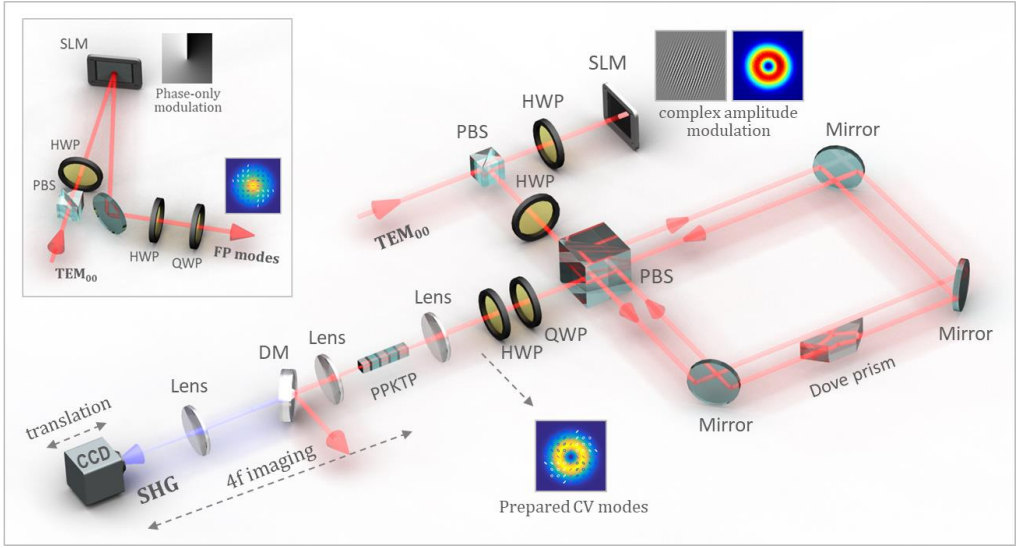

FIG. 7. Schematic setup to implement vectorial type-II SHG with CV-mode pump fields. The upper left inset shows the setup used to generate FP-mode pump fields; see text for details. The key components include the polarizing beam splitter (PBS), half wave plate (HWP), quarter wave plate (QWP), spatial light modulator (SLM), and dichroic mirror (DM).

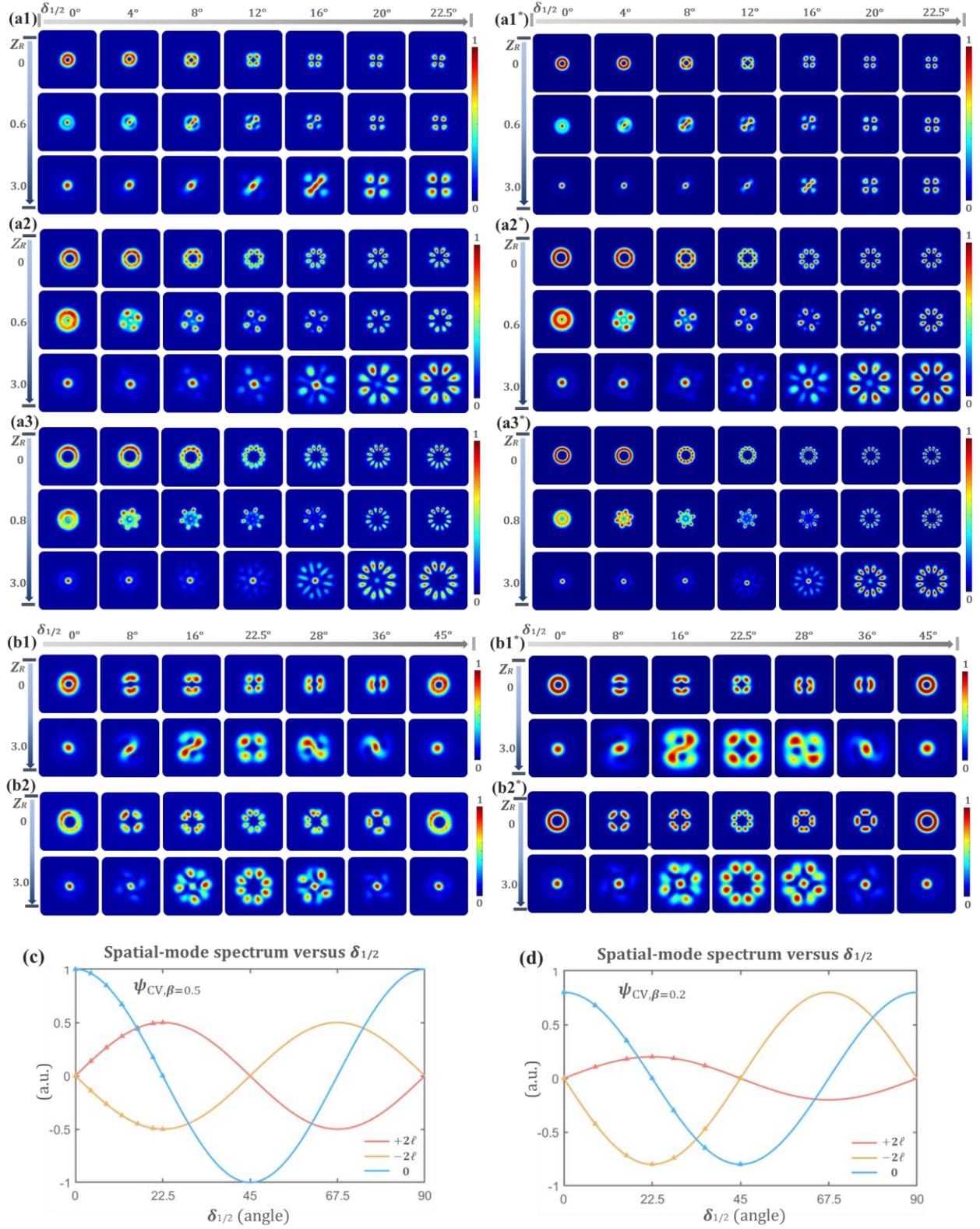

FIG. 8. Observed SHG beam profiles (eight-bit false-color greyscale) driven by CV-mode pump fields from $z_0$ to $z_R = 3$ for different angles $\delta_{l/2}$, where (a1)–(a3) correspond to $\ell = 1, 2$ and $3$ and $\beta=0.5$, and (b1)–(b2) correspond to $\ell = 1, 2$ and $\beta=0.2$. (a1*)–(b2*) shown in the right column present simulated observables for comparison (with the same false-color greyscale). (c) and (d) show the theoretical spatial mode spectra of the corresponding SHG fields as a function of $\delta_{l/2}$.

We then demonstrate predictions with respect to the SHG pumped by the FP modes, where the FP modes with $\beta = 0.5$ and $\ell = 1, 2$ were considered. Figures 9 (a) and (b) show the experimental observations and corresponding theoretical results, where we see that the observed beam profiles from the NP plane to the far field agree well with the theory. As discussed in Sec. III (B), we also confirmed that the far-field beam profiles, except for $\delta_{1/2} = 0°, 45°$, all underwent a 90° rotation with respect to the generation plane. Moreover, the theoretical spatial mode spectra of the observed SHG fields as a function of $\delta_{1/2}$ are shown in Figs. 8 (e) and (f). Note that, as mentioned in Sec. III (B), the spatial mode spectrum changed with the $\ell$ carried by the pump.

We also demonstrated the prediction with respect to the influence of the pump SOC on the intensity of the NP. We chose four typical predictions discussed in Sec. III, i.e., $I_{NP}$ as function of $\delta_{1/2}$ and $\delta_{1/4}$ shown in Figs. 3 (b) and (c); $I_{NP}$ as a function of $\delta_{1/2}$ for $\beta = 0 \sim 0.5$ as shown in Fig. 4 (b), and $I_{NP}$ as a function of $\delta_{1/2}$ shown in Fig. 6 (a). The experimental observations shown in Fig. 10 match perfectly with the theory once again.

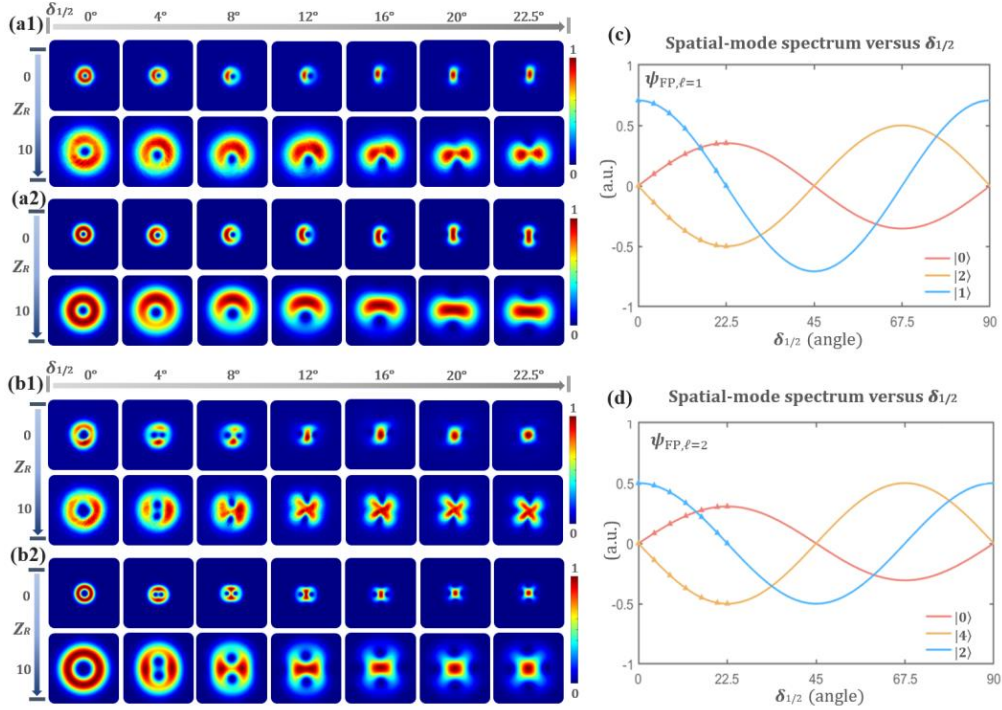

FIG. 9. Observed SHG beam profiles (eight-bit false-color greyscale), driven by the FP-mode pump fields, from $z_0$ to $z_R = 10$, versus $\delta_{1/2}$, where (a1) and (b1) correspond to the pump fields carrying $\ell = 1$ and $\ell = 2$, respectively, and (a2) and (b2) present the simulated observables for comparison (with the same false-color greyscale). (c) and (d) are the theoretical spatial mode spectra of the corresponding SHG fields as a function of $\delta_{1/2}$.

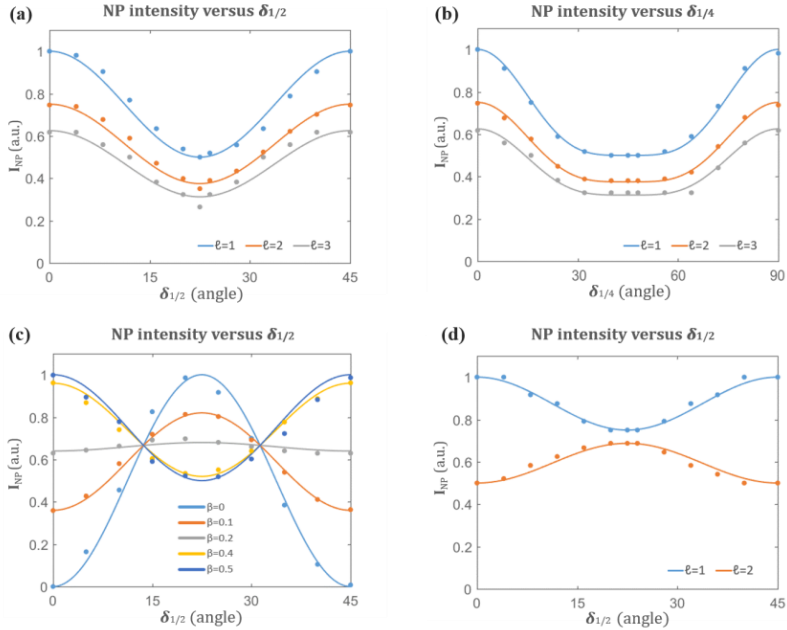

FIG. 10. Observed SHG $I_{NP}$ with respect to the SOC structures of the pump fields, where the point data represent experimental observations and the solid lines represent theoretical data. Plots shown in (a) and (b) correspond to $I_{NP}$ driven by CV-mode pumps, for $\ell = 1, 2, 3$ and $\beta = 0.5$, as a function of $\delta_{1/2}$ and $\delta_{1/4}$. Plots shown in (c) correspond to $I_{NP}$ driven by the CV-mode pumps, for $\ell = 1$ and $\beta = 0 \sim 0.5$, as a function of $\delta_{1/2}$. Plots shown in (d) correspond to $I_{NP}$ driven by the FP-mode pumps for $\ell = 1, 2$ and $\beta = 0.5$ as a function of $\delta_{1/2}$.

## V. DISCUSSION AND CONCLUSION

In this first of the series studies, the results show that although the SHG, as the first nonlinear optical process ever found, has been studied for almost 60 years and is widely used in laser frequency-doubling techniques, the interaction still continues to unfold unexpected outcomes when involving a vector applied laser field. The results here provide a unified description of type-II SHG compatible with both the scalar and the vector cases. The theory can be used, on the one hand, to explain the spatial structures of SHG beams observed in past work [36-44, 55]. On the other hand, it contributes to our fundamental understanding of nonlinear optics mediated by photonic SOC and lays a foundation for future studies, such as on the frequency conversion of SOC states and the generation of vector modes via type-II SHGs.

From Eqs. (11) and (18) as well as the spatial mode spectra shown in Figs. (9) and (10), it is clear that for the SHG pumped by a more general laser mode carrying a net OAM, the OAM selection rule might not be expressed simply as "the OAM of the SHG is double that of input pump." We will discuss this in detail in a separate paper in this series. Note that there is no doubt for OAM conservation in nonlinear optical interactions, however, the OAM selection rule is not constant for a given interaction, such as the abnormal selection rule reported in SBS and SRS [59, 60].

In summary, we presented a general theoretical toolkit for analyzing the type-II SHG driven by the vector laser mode. Based on this, a detailed examination of the SHG driven by typical CV modes and FP modes was provided, where we theoretically revealed and experimentally demonstrated how the SOC structures of pump fields affect and control the intensity and spatial structure of the created NPs. As a consequence, we showed how the beam profiles of the generated SHG fields evolve upon propagation.

## ACKNOWLEDGMENT

This work was supported by the National Natural Science Foundation of China (Grant Nos. 11934013 and 61975047). All the authors acknowledge useful inputs from two professional reviewers.

## Appendix A

The wavefunction of Laguerre-Gaussian (LG) mode with the radial index $p=0$ in cylindrical coordinates $\{r,\varphi,z\}$, used in simulations, is given by [61-63]

$$LG_\ell(r,\varphi,z) = \sqrt{\frac{2}{\pi(|\ell|)!}}\frac{1}{w(z)}\left(\frac{\sqrt{2}r}{w(z)}\right)^{|\ell|}L^{|\ell|}\left(\frac{2r^2}{w^2(z)}\right)$$
$$\times \exp\left(\frac{-r^2}{w^2(z)}\right)\exp[-i\Phi(r,\varphi,z)], \quad (A1)$$

where $\Phi(r,\varphi,z) = kz + \omega r^2/2cR(z) + \ell\varphi - (|\ell|+1)\tan^{-1}(z/z_R)$, $\ell$ is the topological charge giving an OAM of $\ell\hbar$ per photon, $L^{|\ell|}$ is the Laguerre polynomial, $R(z)$ is the curvature radius of the wavefront, $z_R$ is the Rayleigh length for a given beam waist $w_0$, $w(z) = w_0(1+z^2/z_R^2)^{-1}$ and $(|\ell|+1)\tan^{-1}(z/z_R)$ describes the beam expanding and the Gouy phase accumulated during the diffraction propagation, respectively.

## Appendix B

The transformation of the HWP and QWP for the SoPs of paraxial beams can be described by Jones matrices with respect to the fast axis angles $\delta_v$, which are given by

$$\mathbf{M_Q} = \begin{pmatrix} i\cos^2(\delta_{1/4}) + \sin^2(\delta_{1/4}) & (i-1)\sin(\delta_{1/4})\cos(\delta_{1/4}) \\ (i-1)\sin(\delta_{1/4})\cos(\delta_{1/4}) & i\sin^2(\delta_{1/4}) + \cos^2(\delta_{1/4}) \end{pmatrix},$$

and

$$\mathbf{M_H} = \begin{pmatrix} \cos(2\delta_{1/2}) & \sin(2\delta_{1/2}) \\ \sin(2\delta_{1/2}) & -\cos(2\delta_{1/2}) \end{pmatrix}, \quad (B1)$$

respectively. Therefore, for a given polarizations $\hat{\mathbf{e}}_{+/-}$, the SoPs after the transformation can be derived from $\mathbf{M}_{H/Q}\hat{\mathbf{e}}_{+/-}$.